\documentclass[12pt]{article}
\usepackage{amssymb}
\usepackage{cite}
\usepackage{rotating}
\usepackage{epsfig}
\newcommand{\be}{\begin{equation}}
\newcommand{\ee}{\end{equation}}
\newcommand{\bd}{\begin{description}}
\newcommand{\ed}{\end{description}}
\newcommand{\bmat}{\begin{displaymath}}
\newcommand{\emat}{\end{displaymath}}
\newcommand{\bit}{\begin{itemize}}
\newcommand{\eit}{\end{itemize}}
\newcommand{\ben}{\begin{enumerate}}
\newcommand{\een}{\end{enumerate}}
\newcommand{\bc}{\begin{center}}
\newcommand{\ec}{\end{center}}
\begin{document}
\bc
{\bf \Large On higher order corrections
to photon structure functions}\ec

\bc Ji\v{r}\'{\i} Ch\'{y}la\\
Institute of Physics of the Academy of Sciences of the Czech Republic\\
Na Slovance 2, 18221 Prague 8, Czech Republic
\ec

\bc {\bf Abstract}\ec
The QCD corrections to photon structure functions are defined in a way
consistent with the factorization scheme invariance. It is shown that the
conventional DIS$_{\gamma}$ factorization scheme does not respect this
invariance and is thus deeply flawed. The origins of the divergent
behavior of photonic coefficient function at large $x$ are analyzed and
recipe to remove it is suggested.

\noindent
\section{Introduction}
\begin{figure}\unitlength=1mm
\begin{picture}(170,60)
\put(30,8)
{
\begin{sideways}
\begin{sideways}
\begin{sideways}
{\epsfig{file=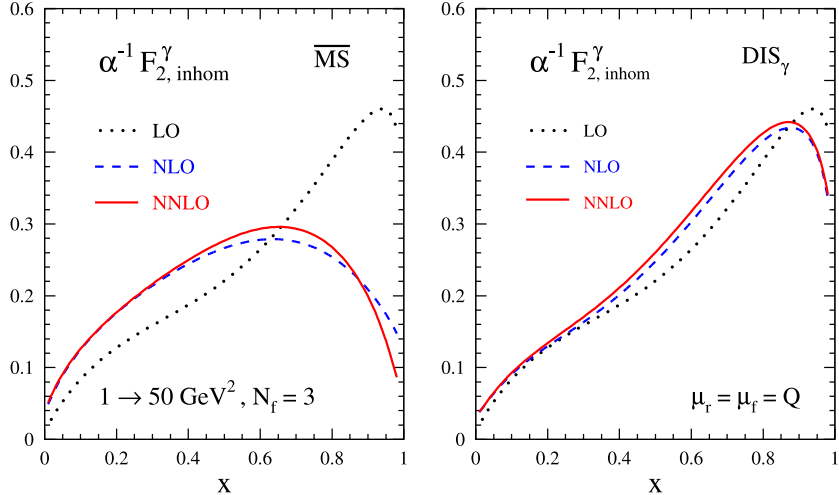,width=3.9cm}}
\end{sideways}
\end{sideways}
\end{sideways}
}
\end{picture}
\caption{The comparison of LO, NLO and NNLO approximations to the
inhomogeneous part of the photon
structure function in the $\overline{\mathrm{MS}}$ and
DIS$_{\gamma}$ FS.
Taken from \cite{VogtWarsaw}.}
\label{f2pdf}
\end{figure}

The recently completed evaluation of order $\alpha_s^3$ parton-parton
\cite{MVVhad1,MVVhad2} and order $\alpha\alpha_s^2$ photon-parton splitting
functions \cite{VogtWarsaw} has confirmed the earlier results \cite{MVVph}
based on the evaluation of first six even moments. For both the quark
distribution functions of the photon and the photon structure function,
the results (for the latter shown in Fig. \ref{f2pdf}) exhibit very small
difference between the next-to-leading (NLO) and next-to-next-to-leading
order (NNLO) approximations defined in the standard way and evaluated in
$\overline{\mathrm{MS}}$ factorization scheme (FS).

This welcome feature stands in sharp contrast to the large difference,
noted already in \cite{GRV92}, between the standard LO and NLO approximations
in the same $\overline{\mathrm{MS}}$ FS. The standard formulation of the NLO
approximation to photon structure function $F_2^{\gamma}(x,Q^2)$ suffers also
from problems in the large $x$ region where it turns negative. In order to
cure these two, related, problems, the so called DIS$_\gamma$ FS has been
proposed in \cite{GRV92b}.

The aim of this note is to show that, first, there is no such sharp
difference between the LO and NLO approximations to $F_2^{\gamma}$ in the
$\overline{\mathrm{MS}}$ FS if these approximations are defined in a way
consistent with the factorisation procedure. Second, I will argue that the
way the DIS$_\gamma$ FS is introduced violates the basic requirement of the
FS invariance and, consequently, the related definition of parton distribution
functions of the photon is deeply flawed. To see where the problem comes from,
I will contrast the definition of DIS$_\gamma$ FS with the analogous, but
theoretically well-defined concept of DIS FS in the case of the
proton structure function.

The main source of the difference between the standard treatment of the proton
and photon structure functions can be traced back to the interpretation of
the behaviour of parton distribution functions of the photon in perturbative
QCD. In \cite{ja} I have discussed this point at length, but its conclusions
have been largely ignored. I will therefore return to this point and present
additional arguments showing that parton distribution functions of the
photon behave like $\alpha$, rather than $\alpha/\alpha_s$ as claimed in
\cite{VogtWarsaw,MVVph,GRV92,GRV92b} and most, though not all, other papers
on photon structure functions. I will outline how the theoretically consistent
LO, NLO and NNLO QCD approximations to $F_2^{\gamma}$ should be constructed
from the results of \cite{VogtWarsaw,MVVph}.

Finally, I will discuss the origins of the problems in the large $x$
region and suggest how to remove them taking into account
their pure QED nature.

\section{Basic facts and notation}
Let us start by briefly recalling the basic facts concerning
the various ingredients of perturbative calculations involving
(quasi)real photons in the initial state.
In QCD the coupling of quarks and gluons is characterized by the
renormalized colour coupling (``couplant'' for short)
$\alpha_s(\mu)$, depending on the {\em renormalization scale} $\mu$
and satisfying the equation
\begin{equation}
\frac{{\mathrm d}\alpha_s(\mu)}{{\mathrm d}\ln \mu^2}\equiv
\beta(\alpha_s(\mu))=
-\frac{\beta_0}{4\pi}\alpha_s^2(\mu)-
\frac{\beta_1}{16\pi^2}
\alpha_s^3(\mu)+\cdots,
\label{RG}
\end{equation}
where, in QCD with $n_f$ massless quark flavours, the first two
coefficients, $\beta_0=11-2n_f/3$ and $\beta_1=102-38n_f/3$, are
unique, while all the higher order ones are ambiguous. However,
even for a given r.h.s. of (\ref{RG}) there is an infinite number
of solutions, differing by the initial condition. This so called {\em
renormalization scheme} (RS) ambiguity
\footnote{In higher orders
this ambiguity includes also the arbitrariness of the coefficients
$\beta_i,i\ge 2$.}
can be parameterized in a number of ways, for instance
by the value of the renormalization scale, usually denoted $\Lambda$,
for which $\alpha_s(\mu=\Lambda_{\mathrm{RS}})=\infty$. At the NLO
the variation of both the renormalization scale $\mu$ and the
renormalization scheme RS$\equiv$\{$\Lambda_{\mathrm {RS}}$\} is
legitimate but redundant. It suffices to fix one of them and vary the
other. In this paper we shall work in the standard
$\overline{\mathrm {MS}}$ RS.

As nothing in my arguments depends in essential way on the numerical
value of $\beta_1$, I will set in the following $\beta_1=0$. This
assumption, simplifies many otherwise complicated formulae and lays bare
the essential aspects of the problem. Note, however, that this assumption
does not amount to working in the leading order of QCD, as all the
relevant terms of higher order QCD approximations and their relations
are kept.

Factorization scale dependence of PDF of the photon
is determined by the system of inhomogeneous evolution equations
\begin{eqnarray}
\frac{{\mathrm d}\Sigma(x,M)}{{\mathrm d}\ln M^2}& =&
k_{\Sigma}+P_{qq}\otimes \Sigma+ 2n_fP_{qG}\otimes G,
\label{Sigmaevolution}
\\ \frac{{\mathrm d}G(x,M)}{{\mathrm d}\ln M^2} & =& k_G+
P_{Gq}\otimes \Sigma+ P_{GG}\otimes G, \label{Gevolution} \\
\frac{{\mathrm d}q_{\mathrm {NS}}(x,M)}{{\mathrm d}\ln M^2}& =&
k_{\mathrm {NS}}+P_{\mathrm {NS}}\otimes q_{\mathrm{NS}},
\label{NSevolution}
\end{eqnarray}
where
\begin{eqnarray}
\Sigma(x,M) & \equiv & \sum_{i=1}^{n_f}q_i^{+}(x,M)\equiv
\sum_{i=1}^{n_f} \left[q_i(x,M)+\overline{q}_i(x,M)\right],
\label{singlet}\\ q_{\mathrm{NS}}(x,M)& \equiv &
\sum_{i=1}^{n_f}\left(e_i^2-\langle e^2\rangle\right)
\left(q_i(x,M)+\overline{q}_i(x,M)\right),
\label{nonsinglet}
\end{eqnarray}
\begin{equation}
k_{\mathrm {NS}}\equiv\delta_{\mathrm{NS}}k_q;~~~
\delta_{\mathrm{NS}}=6n_f\left(\langle e^4\rangle-\langle
e^2\rangle ^2\right),~~~k_{\Sigma}\equiv\delta_{\Sigma}k_q;~~~
\delta_{\Sigma}=6n_f\langle e^2\rangle.
\label{sigmas}
\end{equation}
The splitting functions $P_{ij}$ and
$k_i$ are given as power expansions in $\alpha_s(M)$:
\begin{eqnarray}
k_i(x,M) & = & \frac{\alpha}{2\pi}\left[k^{(0)}_i(x)+
\frac{\alpha_s(M)}{2\pi}k_i^{(1)}(x)+
\left(\frac{\alpha_s(M)}{2\pi}\right)^2k^{(2)}_i(x)+\cdots\right],
\label{splitquark} \\
P_{ij}(x,M) & = &
~~~~~~~~~~~~~~~~~~\frac{\alpha_s(M)}{2\pi}P^{(0)}_{ij}(x) +
\left(\frac{\alpha_s(M)}{2\pi}\right)^2 P_{ij}^{(1)}(x)+\cdots,
\label{splitpij}
\end{eqnarray}
where $i=q,G,{\mathrm{NS}}$ and $ij=qq,qG,Gq,GG,{\mathrm{NS}}$.
The leading order splitting functions
$k_q^{(0)}(x)$ and $P^{(0)}_{ij}(x)$ are {\em
unique}, while all higher order ones
$k^{(j)}_q,k^{(j)}_G,P^{(j)}_{kl},j\ge 1$ depend on the choice of
the {\em factorization scheme} (FS).
According to the factorization theorem, the photon structure
function $F_2^{\gamma}(x,Q^2)$ is given as
the sum of the convolutions
\begin{eqnarray}
\frac{1}{x}F_2^{\gamma}(x,Q^2)&=& q_{\mathrm{NS}}(M)\otimes
C_q(Q/M)+\delta_{\mathrm{NS}}C_{\gamma}+
\label{NSpart} \\ & & \langle e^2\rangle \Sigma(M)\otimes
C_q(Q/M)+ \langle
e^2\rangle\delta_{\Sigma}C_{\gamma}+ \langle
e^2\rangle G(M)\otimes C_G(Q/M)
\label{S+Gpart}
\end{eqnarray}
of PDF and coefficient functions
$C_q,C_G,C_{\gamma}$ admitting perturbative expansions
\begin{eqnarray}
C_q(x,Q/M) & = & \delta(1-x)~~~~~~~~~~~~~~+
\frac{\alpha_s(\mu)}{2\pi}C^{(1)}_q(x, Q/M)+\cdots,
\label{cq} \\
C_G(x,Q/M) & = & ~~~~~~~~~~~~~~~~~~~~~~~~~~~~
\frac{\alpha_s(\mu)}{2\pi}C^{(1)}_G(x,Q/M)+\cdots,
\label{cG} \\
C_{\gamma}(x,Q/M) & = &\frac{\alpha}{2\pi}\left[
C_{\gamma}^{(0)}(x,Q/M)+
\frac{\alpha_s(\mu)}{2\pi}C_{\gamma}^{(1)}(x,Q/M)+\cdots\right],
\label{cg}
\end{eqnarray}
where the standard formula for $C_{\gamma}^{(0)}$ reads
\begin{equation}
C_{\gamma}^{(0)}(x,Q/M)=
\left(x^2+(1-x)^2\right)\ln\frac{Q^2(1-x)}{M^2x}+8x(1-x)-1.
\label{C0}
\end{equation}
I am using the terms ``renormalization'' and ``factorization'' scales in
standard way as discussed, for instance, in \cite{catani,collins,soper}.
However, some authors use these concepts in a very different way. For
instance, in \cite{ewald} the renormalization scale denotes the argument of
the couplant $\alpha_s$ in the expansion (\ref{splitpij}) of the splitting
functions, rather than in the expansions (\ref{cq}-\ref{cg}) of the
coefficient functions as in \cite{catani,collins,soper} and the present
paper. This alternative definition of the renormalization scale, though
mathematically legitimate, lacks physical motivation as it fails to make
the crucial difference, emphasized long time ago by Politzer \cite{politzer},
between the ambiguities of the treatment of perturbatively calculable short
distance physics, embodied in the renormalization scale and described by the
coefficient functions, and those of the large distance, nonperturbative
effects that go into the PDF and induce their dependence on $M$.

The renormalization scale $\mu$, used as argument of $\alpha_s(\mu)$
in (\ref{cq}-\ref{cg}) is in principle independent of the
factorization scale $M$. Note that despite the presence of $\alpha_s(\mu)$
in (\ref{cq}--\ref{cg}), the coefficient functions $C_q,C_G$ and $C_{\gamma}$
are, if calculated to all orders in $\alpha_s$, independent of $\mu$
as well as of the RS. On the
other hand, PDF and the coefficient functions $C_q, C_G$ and
$C_{\gamma}$ do depend on both the factorization scale $M$ and
factorization scheme, but in such a correlated manner that
physical quantities, like $F_2^{\gamma}$, are independent of both $M$
and the FS, provided expansions (\ref{splitquark}--\ref{splitpij})
and (\ref{cq}--\ref{cg}) are taken to all orders. In practical calculations
based on truncated forms
of (\ref{splitquark}--\ref{splitpij}) and (\ref{cq}--\ref{cg}) this
invariance is, however, lost and the choice of both $M$ and FS
makes numerical difference even for physical quantities.

\section{Parton distribution functions of the photon}
The general solution of the evolution equations
(\ref{Sigmaevolution}-\ref{NSevolution}) can be written as the sum
of a particular solution of the full inhomogeneous equation and the
general solution of the corresponding homogeneous one, called
{\em hadronic}. In rest of this note I will
for technical reasons restrict the discussion to
the nonsinglet quark distribution function of the photon and the
corresponding nonsinglet part (\ref{NSpart}) of photon structure
function. To simplify the formulae I will also
drop the subscript ``NS'' and set $\delta_{\mathrm{NS}}=1$ everywhere.

\begin{figure}
\epsfig{file=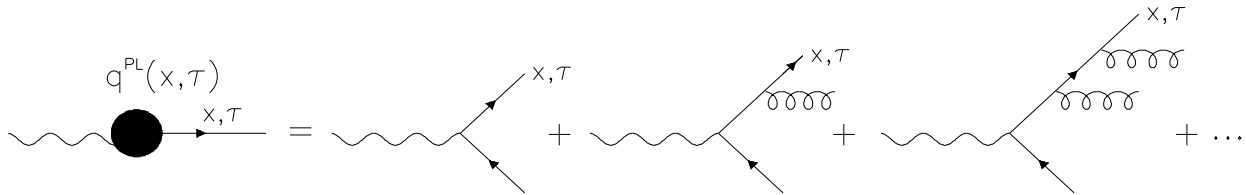,width=\textwidth}
\caption{Diagrams
defining the pointlike part of non-singlet quark distribution
function of the
photon in the leading logarithmic approximation.}
\label{figpl}
\end{figure}

As is well known, the subset of the solutions of the
evolution equation (\ref{NSevolution}) with the splitting functions
including the first terms $k^{(0)}$ and $P^{(0)}$ only and
vanishing at some scale $M_0$ results from the resummation of the
contributions of the diagrams in Fig. (\ref{figpl}). These, so called
{\em pointlike} solutions, which start with the purely QED vertex
$\gamma\rightarrow q\overline{q}$, define the standard
``leading order'' approximation and have, in momentum space, the form
(again I will drop the specification ``pointlike'' throughout the rest
of this paper)
\begin{equation}
q(n,M_0,M)=\frac{4\pi}{\alpha_s(M)}a(n)
\left[1-\left(\frac{\alpha_s(M)}{\alpha_s(M_0)}\right)^
{1-2P^{(0)}(n)/\beta_0}\right],
\label{generalpointlike}
\end{equation}
where
\begin{equation}
a(n)\equiv \frac{\alpha}{2\pi\beta_0}
\frac{k^{(0)}(n)}{1-2P^{(0)}(n)/\beta_0}.
\label{ans}
\end{equation}
As argued in \cite{ja} the fact that $\alpha_s(M)$ appears in the
denominator of (\ref{generalpointlike}) does in no way mean that
$q\propto \alpha/\alpha_s$. It is obvious
that provided $M$ and $M_0$ are kept fixed when QCD is switched off by
sending $\Lambda\rightarrow 0$ the expression (\ref{generalpointlike})
approaches
\begin{equation}
q(x,M,M_0)\longrightarrow
\frac{\alpha}{2\pi} k^{(0)}(x)
\ln\frac{M^2}{M_0^2}\, ,
\label{QEDlimit}
\end{equation}
corresponding to purely QED splitting $\gamma\rightarrow q\overline{q}$.
Note that this limit holds even if we include in (\ref{NSevolution}) the
term proportional to $k^{(1)}$, which gives the lowest order QCD
contribution to the inhomogeneous splitting function $k_q$.

To see what is wrong with the conventional way of arriving at the claim
that $q\propto \alpha/\alpha_s$ let us recast the evolution
equation (\ref{NSevolution}) into the equivalent form
\be
\frac{\mathrm{d}q(n,Q)}{\mathrm{d}\alpha_s}=
-\frac{4\pi}{\beta_0}\left[\frac{\alpha}{2\pi}\frac{k^{(0)}(n)}{\alpha_s^2}
+\frac{\alpha}{2\pi}\frac{P^{(0)}(n)}{\alpha_s}q(n,Q)
+\cdots\right]
\label{NSequiv}
\ee
and take into account just the first term on its the right. Trivial
integration then yields
\be
q(n,Q)=
\frac{\alpha}{2\pi}\frac{4\pi}{\beta_0}\frac{k^{(0)}(n)}{\alpha_s(Q)}+A,
\label{sol}
\ee
where $A$ denotes arbitrary integration constant specifying the boundary
condition on the solution of (\ref{NSequiv}). Choosing $A=0$ might, but
should not, mislead us to the usual claim that $q\propto 1/\alpha_s$, because
(\ref{NSequiv}) is equivalent to (\ref{NSevolution}), which in our
approximation of keeping just the first term
on its r.h.s. contains no trace of QCD, being of purely QED nature! Taking
the difference of (\ref{sol}) for $Q_1$ and $Q_2$ and inserting the explicit
expression for $\alpha_s(M)$ we arrive at
\be
q(n,Q_1)-q(n,Q_2)=
\frac{\alpha}{2\pi}\frac{4\pi}{\beta_0}k^{(0)}(n)\left(
\frac{1}{\alpha_s(Q^2_1)}-\frac{1}{\alpha_s(Q^2_2)}\right)
=\frac{\alpha}{2\pi}k^{(0)}(n)\ln\frac{Q^2_1}{Q_2^2},
\label{man1}
\ee
the purely QED expression we can get directly from (\ref{NSevolution}).
We can further recast (\ref{man1}) into the form similar to
(\ref{generalpointlike})
\be
q(n,Q_1)-q(n,Q_2)=
\frac{4\pi}{\alpha_s(Q_1^2)}\frac{\alpha}{2\pi\beta_0}
\frac{k^{(0)}(n)}{d}
\left[1-\left(\frac{\alpha_s(Q_1^2)}{\alpha_s(Q^2_2)}\right)^d\right]
\label{man2}
\ee
with $d=1$. Including also the lowest order QCD term proportional to
$P^{(0)}(n)$ modifies the result slightly, but non-essentially by
replacing $d=1$ with $d=1-2P^{(0)}(n)/\beta_0$. Note that since
$P^{(0)}_{qq}(0)=0$ the integral of quark distribution function is
actually unchanged by QCD effects, but the
standard claim would still be that it ``behaves'' as $1/\alpha_s$.

By the same reasoning we could ``prove'' that, for instance, also
the vacuum polarization $\Pi(Q^2)$ ``behaves like $1/\alpha_s$''!
Indeed, taking the derivative of $\Pi(Q^2)$ with respect to $\ln Q$
\be
D(Q^2)\equiv
-\frac{\mathrm{d}\Pi(Q^2)}{\mathrm{d}\ln Q^2}=
-\frac{\mathrm{d}\Pi(Q^2)}{\mathrm{d}\alpha_s(Q)}\frac{\mathrm{d}\alpha_s(Q)}
{\mathrm{d}\ln Q^2}
\label{adler}
\ee
which defines the Adler function, and rewriting it in the way analogous to
(\ref{NSequiv}) we get
\be
\frac{\mathrm{d}\Pi(Q)}{\mathrm{d}\alpha_s(Q)}=
-\frac{D(Q^2)}{\mathrm{d}\alpha_s(Q)
/{\mathrm{d}\ln Q^2}}=\frac{4\pi}{\beta_0}
\frac{1+\sum_{k=1}^{\infty}d_k\alpha_s^k(Q)}{\alpha_s^2+\cdots}.
\label{adler2}
\ee
Trivial integration then yields the advertised conclusion
\be
\Pi(Q^2)=-\frac{4\pi}{\beta_0}\frac{1}{\alpha_s(Q)}+\cdots.
\label{adler3}
\ee
Let me finally add another, quite different, argument demonstrating
that PDF of the photon behave as ${\cal O}(\alpha)$. Consider the
Mellin moments of the nonsinglet part of
$F^{\gamma}(x)\equiv F_{\mathrm{NS}}^{\gamma}(x,Q)/x$
\begin{equation}
F^{\gamma}(n,Q)=q(n,M) C_q(n,Q/M)+C_{\gamma}(n,Q/M)
\label{FNS}
\end{equation}
for the general pointlike solution of (\ref{NSevolution}), which
like (\ref{generalpointlike}) vanishes at $M=M_0$.
As $F^{\gamma}(n,Q)$ is independent of the factorization scale
$M$, we can take any $M$ to evaluate it, for instance just
$M_0$. However, for $M=M_0$ the first term in (\ref{FNS}) vanishes
and we get
\footnote{In the rest of this paper the dependence on the Mellin moment
variable $n$ will be suppressed.}
\begin{equation}
F^{\gamma}(Q)=\frac{\alpha}{2\pi}\left[C_{\gamma}^{(0)}(Q/M_0)+
\frac{\alpha_s(\mu)}{2\pi}C_{\gamma}^{(1)}(Q/M_0)+
\left(\frac{\alpha_s(\mu)}{2\pi}\right)^2C_{\gamma}^{(2)}(Q/M_0,Q/\mu)
+\cdots\right]
\label{cgammaonly}
\end{equation}
i.e. manifestly the expansion in powers of
$\alpha_s(\mu)$ which starts with ${\cal O}(\alpha)$ pure QED
contribution $(\alpha/2\pi)C_{\gamma}^{(0)}$ and includes standard
QCD corrections of orders $\alpha_s^k,k\ge 1$. Clearly this expansions
vanishes when QCD is switched off and there is no trace of the supposed
``$\alpha/\alpha_s$'' behaviour. I will come back to this expression
in Section 4.

\section{Defining LO and NLO approximations
for ${\mathbf{F^{\gamma}(x,Q^2)}}$}
Although semantics is a matter of convention, I think it is wise to
define the terms ``leading'', ``next--to--leading'' and higher orders
in a way which guarantees that they have the same meaning in different
processes. Recall that for the case of the familiar ratio
\begin{equation}
R_{\mathrm{e}^+\mathrm{e}^-}(Q)\equiv
\frac{\sigma(\mathrm{e}^+\mathrm{e}^-\rightarrow
\mathrm{hadrons})}{\sigma(\mathrm{e}^+\mathrm{e}^-\rightarrow
\mu^+\mu^-)}
=\left(3\sum_{i=1}^{n_f}e_i^2\right)\left(1+r(Q)\right)
\label{Rlarge}
\end{equation}
the prefactor $3\sum_{i=1}^{n_f}e_i^2$, which comes from pure QED, is
usually subtracted and only the QCD correction $r(Q)$
is considered for further analysis. For the quantity (\ref{Rlarge}) the
terms ``leading'' and ``next--to--leading'' thus apply only to genuine
QCD effects as described by $r(Q)$, which starts as $\alpha_s/\pi$.
Unfortunately, this practice is ignored in most analyses
of $F^{\gamma}$, which count the QED term $k^{(0)}$ as the ``leading order''
\cite{VogtWarsaw,MVVph,GRV92,GRV92b}.

In \cite{ja} I have proposed the definition
of QCD approximations of $F^{\gamma}$ which follows closely the convention
used in QCD analysis of quantities like (\ref{Rlarge}).
It starts with writing the (pointlike nonsinglet) quark distribution
function $q(M)$ as the sum of the purely QED contribution
\begin{equation}
q_{\mathrm{QED}}(M)\equiv\frac{\alpha}{2\pi}k^{(0)}
\ln\frac{M^2}{M_0^2}
\label{qQED}
\end{equation}
and the QCD correction satisfying the inhomogeneous evolution
equation
\begin{eqnarray}
\frac{\mathrm{d}q_{\mathrm{QCD}}(M)}{{\mathrm{d}}\ln M^2}& = &
\frac{\alpha_s}{2\pi}\left[\frac{\alpha}{2\pi}k^{(1)}+P^{(0)}
q_{\mathrm{QED}}(M)\right]+
\left(\frac{\alpha_s}{2\pi}\right)^2\left[
\frac{\alpha}{2\pi}k^{(2)}+P^{(1)}q_{\mathrm{QED}}(M)\right]
+\cdots  \label{inh}\nonumber\\
& +& \frac{\alpha_s}{2\pi}P^{(0)}q_{\mathrm{QCD}}(M)+
\left(\frac{\alpha_s}{2\pi}\right)^2P^{(1)}q_{\mathrm{QCD}}(M)+\cdots.
\label{modified}
\end{eqnarray}
The latter differs from that satisfied by the full quark distribution
function not only by the absence of the term
$(\alpha/2\pi)k^{(0)}$ but also by shifted appearance of higher
order coefficients $k^{(i)};i\ge 1$. For instance, the
inhomogeneous splitting function $k^{(1)}_q$ enters (\ref{modified})
 at the same order as homogeneous
splitting function $P^{(0)}$ and thus these splitting
functions will appear at the same order also in its solutions.
Similarly, the simultaneous presence of
$k^{(2)}_q$ and $P^{(1)}$ in the $O(\alpha_s^2)$ term of the
inhomogeneous part of (\ref{modified}) implies that the
NLO QCD analysis of $F^{\gamma}$ requires the knowledge of
$k^{(2)}$ etc.
In Table \ref{orders} the terms included in the standard definition
of the LO and NLO approximations are compared with those corresponding
to my definition of these approximations. The difference between the
two definitions is substantial.
\begin{table}[h]
\begin{center}
\begin{tabular}{|l|l|l|}
\hline  \hline   & standard definition& my definition  \\ \hline
QED & does not introduce & $k^{(0)}$, $C_{\gamma}^{(0)}$ \\ \hline
  LO QCD& $k^{(0)}$, $P^{(0)}$ & $k^{(0)}$, $C_{\gamma}^{(0)}$,
  $k^{(1)}$, $C_{\gamma}^{(1)}$,
  $P^{(0)}$
    \\ \hline
 NLO QCD & $k^{(0)}$, $P^{(0)}$, $k^{(1)}$, $C_{\gamma}^{(0)}$,
   $P^{(1)}$ & $k^{(0)}$, $C_{\gamma}^{(0)}$,
  $k^{(1)}$, $C_{\gamma}^{(1)}$, $P^{(0)}$, $k^{(2)}$,
   $C_{\gamma}^{(2)}$,
   $P^{(1)}$ \\
\hline \hline
\end{tabular}
\end{center}
\caption{Contributions included in the standard and correct
definition of LO and NLO.}
\label{orders}
\end{table}

\section{Factorization schemes and their choice - the proton}
Let us first recall the definition of factorization schemes
in the case of the (nonsinglet) proton structure function. Denoting
$F^p\equiv F^p_{\mathrm{NS}}/x$ we can write its moments as the product
\be
F^p(Q)= q(M)C_q(Q/M)
\label{facproton}
\ee
of the coefficient functions $C_q(Q/M)$, given in (\ref{cq}), and the
 quark distribution function
$q(M)$ of the proton. If both the homogeneous splitting function
(\ref{splitpij}) and coefficient function (\ref{cq}) are calculated to all
orders in $\alpha_s$, the product on the r.h.s. of (\ref{facproton}) is
independent of both the factorization scale $M$ and the factorization scheme.
It must thus hold
\be
\frac{{\mathrm{d}}F(Q)}{{\mathrm{d}}\ln M^2}=
q(M)\left[P(M)C_q(Q/M)+\dot{C}_q(Q/M)\right]=
\frac{{\mathrm{d}}F(Q)}{{\mathrm{d}}C_q^{(j)}}=0
\label{der}
\ee
If, however, these expansions are truncated, the resulting
finite order approximations for $F(Q)$ will depend on both the factorization
scale and scheme. For a finite order approximation to be theoretically
consistent, its dependence on all free parameters must
be formally of higher order in $\alpha_s$ than those included in the
approximation. From (\ref{der}) we get, denoting
$\dot{f}(M)\equiv {\mathrm{d}}f(M)/{\mathrm{d}}\ln M^2$,
\begin{equation}
P(M)C_q(Q/M)+\dot{C}_q(Q/M)=0
\label{dFp}
\end{equation}
and expanding the splitting and coefficient functions to
order $\alpha_s$ we arrive at the relation
\be
\dot{C}_q^{(1)}(Q/M)=-P^{(0)}~~~\Rightarrow ~~~C_q^{(1)}(Q/M)=P^{(0)}
\ln(Q^2/M^2)+C_q^{(1)}(1).
\label{cqeq}
\ee
Taking into account the first two terms in (\ref{splitpij}) the solution
of (\ref{NSevolution}) reads
\be
q(M)=
A\left(\alpha_s(M)\right)^{\displaystyle \frac{-2P^{(0)}}{\beta_0}}
\exp\left( -\frac{2P^{(1)}}{\beta_0}\frac{\alpha_s(M)}{2\pi}\right),
\label{qnsproton}
\ee
where the ($n$-dependent) constant $A$ specifies the boundary condition
\footnote{We can also
use the above expression for $M$ and $M_0$ and by forming their ratio
obtain instead of (\ref{qnsproton}) the usual relation between $q(M)$ and
$q(M_0)$.}.
Inserting (\ref{qnsproton}) into (\ref{facproton}) yields
\be
F^p(Q)=
A \left(\alpha_s(M)\right)^{\displaystyle \frac{-2P^{(0)}}{\beta_0}}
\exp\left( -\frac{2P^{(1)}}{\beta_0}\frac{\alpha_s(M)}{2\pi}\right)
\left[1+\frac{\alpha_s(\mu)}{2\pi}C_q^{(1)}(Q/M)\right]
\label{allproton}
\ee
and expanding the exponential we get
\be
F^p(Q)\doteq
A \left(\alpha_s(M)\right)^{\displaystyle \frac{-2P^{(0)}}{\beta_0}}
\left[1+\frac{\alpha_s(\mu)}{2\pi}\left(C_q^{(1)}(Q/M)
-\frac{2P^{(1)}}{\beta_0}\right)\right].
\label{allproton2}
\ee
FS invariance of (\ref{allproton}) implies that
the non-universal functions $C_q^{(1)}$ and $P_{qq}^{(1)}$ are related
\be
C_q^{(1)}(1)=\frac{2P^{(1)}}{\beta_0}+ \kappa~~\Leftrightarrow
P^{(1)}=\frac{\beta_0}{2}\left(C_q^{(1)}(1)-\kappa\right),
\label{cqform}
\ee
where the quantity $\kappa=\kappa(n)$ is
\underline{factorization scheme invariant}.
In other words either $P^{(1)}$ or $C_q^{(1)}$, but not both independently,
can be chosen at will to specify the FS. For instance, $F^p(Q)$ can be written
as a function of $C_q^{(1)}$ explicitly as
\be
F^p(Q)=
A \left(\alpha_s(M)\right)^{\displaystyle \frac{-2P^{(0)}}{\beta_0}}
\exp\left[-\frac{\alpha_s(M)}{2\pi}
\left(C_q^{(1)}-\kappa
\right)\right]
\left(1+\frac{\alpha_s(\mu)}{2\pi}C_q^{(1)}(Q/M)\right).
\label{allproton3}
\ee
The above expression is independent of $C_q^{(1)}$ to the order considered but
not exactly and thus its numerical value does depend on the choice of $C_q^{(1)}$.
Two points are worth noting.

First, the cancelation mechanism based on the relation (\ref{cqform})
operates independently of the value of $\alpha_s$ as well as for any fixed
boundary condition specified by the constant $A=A(n)$.
These constants, being determined by the behaviour of $q(M)$ at asymptotic
values of $M$, provide unambiguous way of specifying the initial condition on
the solution of the evolution equations. The same boundary condition must
therefore be used in all FS.
If the boundary condition on $q(M)$ is specified by the value $q(M_0)$ at some
initial $M_0$, the situation is slightly more complicated. Imagine, we have
$q(M,M_0)$ given as the solution of (\ref{NSevolution}) in a FS specified by
$P^{(1)}$, or equivalently $C_q^{(1)}$, and with the boundary condition given
by $q(M_0)$ at some $M_0$. To get the boundary condition in a FS specified by
$\overline{P}^{(1)}$, or equivalently $\overline{C}_q^{(1)}$, which would yield
the same proton structure function (\ref{facproton}) we first use
(\ref{qnsproton}) to convert the information on $q(M_0)$ into the knowledge of
the constants $A=A(n)$. Once we have them we can again use (\ref{qnsproton})
to compute the appropriate boundary condition $\overline{q}(M_0)$ in the new FS:
\be
\overline{q}(M_0)=q(M_0)
\exp\left[ \frac{2\left(P^{(1)}-\overline{P}^{(1)}\right)}
{\beta_0}\frac{\alpha_s(M_0)}{2\pi}\right]=q(M_0)
\exp\left[\left(C^{(1)}-\overline{C}^{(1)}\right)
\frac{\alpha_s(M_0)}{2\pi}\right].
\label{trans}
\ee
This boundary condition is different from the one in the original FS at the same
$M_0$, and so would also be the solution $q(M,M_0)$, but after inserting into
(\ref{facproton}) we get the same $F^p(Q)$ since the change of $q(M,M_0)$ is
compensated by the accompanying change of $C_q(Q/M)$. If we perform an analysis
of experimental
data on $F^p(Q)$ by fitting the parameters specifying the initial $q(M_0)$
(and $\Lambda_{\mathrm{QCD}}$) we can thus work in any FS and choose in
principle any initial $M_0$ and get the
same results
\footnote{The independence of the chosen FS, as well as of the value of $M_0$
does, however, requires sufficiently flexible form of the parametrization
of $q(M_0)$.} for $F^p(Q)$.

Second, the expression
(\ref{cqform}) connects quantities, $C_q^{(1)}$ and $P^{(1)}$,
which come from evaluation of Feynman diagrams at different fixed orders: the
former at ${\cal O}(\alpha_s)$ order whereas the latter at ${\cal O}(\alpha_s^2)$.
This is due to the fact that the leading-logarithmic terms proportional to powers
of $P^{(0)}$ are resummed to all orders thereby generating the first term on the
r.h.s. of  (\ref{allproton3}). Only after this resummation does $C_q^{(1)}$
contribute to $F^p$ at the same order as $P^{(1)}$.

In the standard $\overline{\mathrm{MS}}$ FS both
$C_q^{(1)}(1)$ and $P^{(1)}$ are nonzero. From the infinity of other possible
schemes only the so called ``DIS'' FS is regularly used. In this scheme one
sets $Q=M$ and $C_q^{(1)}=0$ in order to keep at the NLO the same
relation between the proton structure function $F(Q)$ and the quark
distribution function $q(Q)$ as at the LO.
In some sense opposite to the DIS FS is the FS $P^{(1)}=0$. In this FS the
evolution equation for quark distribution function has the same form at the
NLO as at the LO. As noted in \cite{jch2} this FS corresponds
\footnote{This holds exactly for $\beta_1=0$, for realistic values of
$\beta_1$ the saddle point is slightly away from $P^{(1)}=0$.}
to the point of local stability \cite{pms} of the
expression (\ref{allproton3}) considered as function of $M, \mu$ and
$C_q^{(1)}$.

\section{Factorization schemes and their choice - the photon}
For the hadronic part of the quark distribution function of the
photon the factorization operates in exactly the same way as for the
proton. Consequently $C_q^{(1)}$ can again be used to label different
factorization schemes and also the relations (\ref{cqform})
do hold.

For the pointlike part of the quark distribution function
of the photon the situation is slightly more complicated as the expression
for photon structure function $F^{\gamma}\equiv F_{\mathrm{NS}}^p/x$ involves
another coefficient function, namely the photonic $C_{\gamma}$
\be
F^{\gamma}(Q)= q(M)C_q(Q/M)+C_{\gamma}(Q/M).
\label{facgamma}
\ee
Consequently, the factorization scale invariance implies
\begin{eqnarray}
\dot{F}^{\gamma}(Q)&=&\dot{q}(M)C_q(Q/M)+q(M)\dot{C}_q(Q/M)+
\dot{C}_{\gamma}(Q/M)=0
\nonumber \\
&=&\left[P(M)C_q(Q/M)+\dot{C}_q(Q/M)\right]q(M)+k(M)C_q(Q/M)
+\dot{C}_{\gamma}(Q/M)
\label{dFgamma}
\end{eqnarray}
The expression in the square bracket vanishes for the same reasons as for
the hadronic structure function and we are thus left with
\begin{eqnarray}
\lefteqn{\dot{F}^{\gamma}(Q)=\frac{\alpha}{2\pi}\left\{
\left[k^{(0)}+\frac{\alpha_s(M)}{2\pi}
k^{(1)}+\cdots\right]
\left[1+\frac{\alpha_s(\mu)}{2\pi}C_q^{(1)}(Q/M)+\cdots\right]+
\right.}
 \nonumber\\&&\left.~~~~~~~~~~~~~~~~~~
\dot{C}^{(0)}_{\gamma}(Q/M)+\frac{\alpha_s(\mu)}{2\pi}
\dot{C}^{(1)}_{\gamma}(Q/M)+\cdots
\right\}=\label{dFgamma2} \\
&& \frac{\alpha}{2\pi}\left\{k^{(0)}+
\dot{C}_{\gamma}^{(0)}(Q/M)
+\frac{\alpha_s}{2\pi}\left[
\dot{C}_{\gamma}^{(1)}(Q/M)
+k^{(1)}+k^{(0)}C_q^{(1)}(Q/M)\right]+\cdots\right\}=0.
\nonumber
\end{eqnarray}
The argument of $\alpha_s$ in the expansion of the coefficient functions
$C_q$ and $C_{\gamma}$, i.e. the renormalization scale $\mu$, is in general
different from the factorization scale $M$. However, I did not write out the
argument of $\alpha_s$ in the last line of (\ref{dFgamma2}) because the
coefficient standing by it is independent of it. The Eq. (\ref{dFgamma2}) implies
\begin{eqnarray}
\dot{C}^{(0)}_{\gamma}(Q/M)& = & -k^{(0)},\label{C0eq}\\
\dot{C}^{(1)}_{\gamma}(Q/M)& = & -k^{(1)}-k^{(0)}C_q^{(1)}(Q/M),\label{C1eq}
\end{eqnarray}
which upon integration and taking into account the relations
(\ref{cqeq}) and (\ref{cqform}) yields
\begin{eqnarray}
C^{(0)}_{\gamma}(Q/M)& =
 & k^{(0)}\ln(Q^2/M^2)+C_{\gamma}^{(0)}(1)\label{C0int}\\
C^{(1)}_{\gamma}(Q/M)& = &
\frac{k^{(0)}P^{(0)}}{2}\ln^2(Q^2/M^2)
+\left(k^{(1)}+k^{(0)}C_q^{(1)}(1)\right)\ln(Q^2/M^2)+C_{\gamma}^{(1)}(1).
\label{C1int}
\end{eqnarray}
Let us now consider the pointlike
solution which results from taking into account beside the pure QED splitting
function $ k^{(0)}$ the lowest order QCD splitting functions
$P^{(0)}$ and $k^{(1)}$:
\begin{equation}
q(M,M_0)=
\frac{4\pi a}{\alpha_s(M)}
\left[1-\left(\frac{\alpha_s(M)}{\alpha_s(M_0)}\right)^{1-2P^{(0)}/\beta_0}
\right]-
\frac{\alpha}{2\pi}
\left[1-\left(\frac{\alpha_s(M)}{\alpha_s(M_0)}\right)^{-2P^{(0)}/\beta_0}
\right]\frac{k^{(1)}}{P^{(0)}},
\label{pointlikeNLO}
\end{equation}
where $a=a(n)$ is defined in (\ref{ans}). This solution satisfies
$q(M_0,M_0)=0$ and for small $\alpha_s$ behaves as
\be
q(M,M_0)\doteq
\frac{\alpha}{2\pi}L_M\left[
k^{(0)}+\frac{\alpha_s(M)}{2\pi}\left(k^{(1)}+\frac{k^{(0)}P^{(0)}}{2}
L_M\right)\right],~L_M\equiv\ln\frac{M^2}{M_0^2}.
\label{pointlimit}
\ee
Multiplying (\ref{pointlimit}) with $C_q$ and adding the lowest two terms
of $C_{\gamma}$ we get
\begin{eqnarray}
\lefteqn{\frac{2\pi}{\alpha}
F^{\gamma}(Q)\doteq
L_M k^{(0)}+C_{\gamma}^{(0)}(Q/M)}
\label{pointf}
\\ & &
+\frac{\alpha_s}{2\pi}\left[C_{\gamma}^{(1)}(Q/M)+
\left(k^{(1)}+k^{(0)}C_q^{(1)}(Q/M)\right)L_M+
\frac{k^{(0)}P^{(0)}}{2}L_M^2\right].
\nonumber
\end{eqnarray}
Replacing $C_{\gamma}^{(1)}$ with (\ref{C1int}) and
$C_q^{(1)}$ with analogous expression (\ref{cqform}), we find that
all terms in the square bracket dependent on $M$ cancel out
\footnote{The analogous cancelation in the pure QED contribution to $F^{\gamma}$
is trivial.}
and we are left with expression which is a function of $Q/M_0$ only.
Moreover, as expected from (\ref{cgammaonly}), it is just
$C_{\gamma}^{(1)}(Q/M_0)$!

Note that the only ambiguity in (\ref{cgammaonly}) comes from the freedom
in the choice of the expansion parameter $\alpha_s(\mu)$, i.e. the choice
of the renormalization scale $\mu$ and renormalization scheme. Because
these ambiguities appear first at order $\alpha_s^2$, the
lowest order QCD correction given by $C_{\gamma}^{(1)}(Q/M_0)$ must
be unique! Consequently, $C_{\gamma}^{(1)}(1)$ as well as the combination
\be k^{(1)}+k^{(0)}C_q^{(1)}(1)=\tilde{\kappa}
\label{k1cq}
\ee
must also be unique, i.e. factorization scheme independent.
The above relation implies that the first non-universal inhomogeneous
splitting function $k^{(1)}$ is, similarly as $P^{(1)}$, a function of
$C_q^{(1)}(1)$ (or the other way around). So $C_q^{(1)}(1)$ and its higher
order partners $C_q^{(i)}$ can be chosen to label the factorization scale
ambiguity also for $F^{\gamma}$.
Once these quantities (which are functions of $n$, or equivalantly, $x$) are
chosen all other free parameters are determined by relation like those in
(\ref{cqform}) or (\ref{k1cq}).

\section{What is wrong with the DIS$_{\gamma}$ factorization scheme?}
The freedom in the definition of quark distribution functions of the photon
related to the non-universality of $C_q^{(j)}, j\ge 1$ is intimately connected
with higher order QCD corrections. The purely QED contribution, given by the sum
of first two terms in (\ref{pointf}), is analogous to QED contribution
$3\sum_{i=1}^{n_f}e_i^2$ in (\ref{Rlarge}) and represents an input into QCD
analysis.
It is manifestly of order $\alpha_s^0=1$, but since in the standard approach
the quark distribution function is - incorrectly as I have argued - assigned
the order $1/\alpha_s$, the lowest order QCD contribution appears to be of
lower order than $C_{\gamma}^{(0)}$. The latter is consequently assigned to the
``next-to-leading'' order \cite{VogtWarsaw,GRV92,GRV92b}
and treated in a similar way as the lowest order QCD
coefficient function $C_q^{(1)}$. This procedure has been motivated in part by the
fact that $C_{\gamma}^{(0)}(x,M)$ as given in (\ref{C0}) turns negative at
large $x$ and even diverges to $-\infty$ when $x\rightarrow 1$.
This has lead the authors of \cite{GRV92} to introduce
{\em ``A factorization scheme avoiding the common perturbative
instability problems encountered in the large-x region.''}
In this so called DIS$_{\gamma}$ factorization scheme
{\em ``... we can use the same boundary conditions for the pointlike LO
and HO distributions}
\be
q^{\gamma}_{\mathrm{PL}}(x,Q^2_0)=
\overline{q}^{\gamma}_{\mathrm{PL}}(x,Q^2_0)=
G^{\gamma}_{\mathrm{PL}}(x,Q^2_0)=0
\label{GRVcit}
\ee
{\em without violating the usual positivity requirements.''} \cite{GRV92}.

Their procedure amounts to of redefining the quark distribution function
of the photon by absorbing the pure QED term
$C_{\gamma}^{(0)}$ in $q(M,M_0)$ according to (see eq. (5) of \cite{GRV92b})
\be
\overline{q}(M,M_0)\equiv q(M,M_0)+\frac{\alpha}{2\pi}C_{\gamma}^{(0)}(1).
\label{subst}
\ee
As this redefinition involves the QED quantity $C_{\gamma}^{(0)}$, and
the notion of quark distribution function inside the photon is well-defined
also in pure QED, the above procedure must make sense even there, without any
QCD effects.
Moreover, as the QCD contribution depends on the numerical value of $\alpha_s$
it cannot cure any problem of the pure QED part.

However, in QED it is straightforward to see that in getting rid off
the troubling $C_{\gamma}^{(0)}$ term the procedure proposed in \cite{GRV92}
violates the requirement of factorization scheme invariance.
Recall that in pure QED the contribution to $F^{\gamma}$
coming from the box diagram regularized by explicit quark mass $m_q$ reads
\cite {witten}
\be
F^{\gamma}_{\mathrm{QED}}(Q)=
\frac{\alpha}{2\pi}C_{\gamma}^{(0)}(Q/m_q)=
\frac{\alpha}{2\pi}\left[k^{(0)}\ln\frac{Q^2}{m_q^2}+
C_{\gamma}^{(0)}(1)\right].
\label{pureQED}
\ee
Introducing the arbitrary scale $M$, we can split it into
\be
q_{\mathrm{QED}}(M)\equiv \frac{\alpha}{2\pi}
k^{(0)} \ln\frac{M^2}{m_q^2}
\label{qQED2}
\ee
interpreted as quark distribution function, and $C_{\gamma}^{(0)}(Q/M)$
given in (\ref{C0}):
\be
F^{\gamma}_{\mathrm{QED}}(Q)=q_{\mathrm{QED}}(M)+
\frac{\alpha}{2\pi}C_{\gamma}^{(0)}(Q/M).
\label{FQED}
\ee
The sum (\ref{FQED}) is manifestly $M$-independent. We can now
redefine $q_{\mathrm{QED}}(M)$ by adding an arbitrary function $f=f(n)$
according to
\be
q_{\mathrm{QED},f}(M)\equiv q_{\mathrm{QED}}(M)+\frac{\alpha}{2\pi}f.
\label{qQED'}
\ee
In order to keep the sum
\be
F^{\gamma}_{\mathrm{QED}}(Q)=q_{\mathrm{QED},f}(M)+
\frac{\alpha}{2\pi}C_{\gamma,f}^{(0)}(Q/M)
\label{FQEDf}
\ee
independent of $M$ and $f$ and equal to (\ref{pureQED})
induces the correlated change of $C_{\gamma}^{(0)}$
\be
C_{\gamma,f}^{(0)}(Q/M)\equiv C_{\gamma}^{(0)}(Q/M)-f.
\label{CQED'}
\ee
Note that the DIS$_{\gamma}$ factorization scheme of \cite{GRV92} corresponds
to $f=C_{\gamma}^{(0)}(1)$. We can write down the evolution equation for
$q_{\mathrm{QED},f}(n,M)$ in the ``$f$-factorization scheme''
\be
\frac{{\mathrm d}q_{\mathrm {QED},f}(M)}{{\mathrm d}\ln M^2}=
\frac{\alpha}{2\pi} k^{(0)},
\label{QEDevol}
\ee
which is the same for all $q_{\mathrm{QED},f}(M)$. What we are, however,
not allowed to do is to use the same boundary condition for all
$q_{\mathrm{QED},f}(M)$, i.e. precisely what has been assumed in (\ref{GRVcit})!
If we do that and impose the boundary condition $q_{\mathrm{QED},f}(M=m_q)=0$,
we get
\be
F^{\gamma}(Q)=\frac{\alpha}{2\pi}\left(C_{\gamma}^{(0)}(Q/m_q)-f\right)
\label{fscheme}
\ee
which depends on the choice of $f$ and only for $f=0$ coincides with the
correct result (\ref{pureQED}). To get the latter we must impose on the
solution of the evolution equation (\ref{QEDevol}) in the $f$-factorization
scheme the appropriate boundary condition:
$q_{\mathrm{QED},f}(M=m_q)=(\alpha/2\pi)f$.
The reason the QED expression (\ref{pureQED}) depends on the FS specified by the
function $f=f(n)$ is clear: there is no all order resummation of the
LL and NLL terms describing the multiple photon emission off the
$q\overline{q}$ that would give rise to terms analogous to the first and
second terms on the r.h.s. of (\ref{qnsproton}).

In QCD the redefinition (\ref{qQED'}) (with $q_{\mathrm{QED}}$ replaced with
full quark distribution function $q(M,M_0)$) induces not only the shift
(\ref{CQED'}) of
$C_{\gamma}^{(0)}$ but also those of $k^{(1)}$ and $C_{\gamma}^{(1)}$
\begin{eqnarray}
k^{(1)}_f&\equiv &k^{(1)}-P^{(0)}f\label{substk}\\
C^{(1)}_{\gamma,f}(Q/M)&\equiv
&C^{(1)}_{\gamma}(Q/M )-C_q^{(1)}(Q/M)f.
\label{substC}
\end{eqnarray}
As in pure QED the DIS$_{\gamma}$ factorization scheme corresponds to
$f=C_{\gamma}^{(0)}(1)$. In \cite{GRV92} and most other NLO
QCD analysis the term proportional to $C_{\gamma}^{(1)}$ is assigned to the NNLO
and thus the above second relation (\ref{substC}) is not written out explicitly.

The above substitutions (\ref{CQED'}-\ref{substC}) are legitimate but
as in QED what we are not allowed to do is to impose on $q_f(M,M_0)$
the same boundary condition for all $f$. If we do that we
straightforwardly find that the photon structure function can be written as
\begin{equation}
F(Q)=\frac{\alpha}{2\pi}\left[\left(C_{\gamma}^{(0)}(Q/M_0)-f\right)+
\frac{\alpha_s(\mu)}{2\pi}
\left(C_{\gamma}^{(1)}(Q/M_0)-C_q^{(1)}(Q/M)f\right)
+\cdots\right]
\label{cgammaonly2}
\end{equation}
which coincides with (\ref{cgammaonly2}) only for $f=0$. Even if we sum
(\ref{cgammaonly2}) to all orders of $\alpha_s$ the result does depend on
the chosen $f$, violating the fundamental requirement of factorization
scheme independence of physical quantities!! As already emphasized, the
QCD corrections
cannot compensate the $f$-dependence of the first pure QED term. In fact
even the all order sum of QCD terms in (\ref{cgammaonly2}) is $f$-dependent.

\section{Removing the problem in large $x$ region}
Because the function $C_{\gamma}^{(0)}(x,Q/M)$ is of pure QED origin, one
might expect it to cause problems also for leptonic structure function of
the photon. This structure function, which had been measured at LEP,
is given by the same expression (\ref{pureQED}) as for quarks,
except for the replacement of $m_q$ with the lepton (muon or electron) mass
$m_l$:
\begin{eqnarray}
\frac{2\pi}{\alpha}F^{\gamma}_{lept}(x,Q)&=&
\left(x^2+(1-x)^2\right)
\ln\frac{Q^2(1-x)}{m_l^2 x}+8x(1-x)-1 \nonumber\\
& =&\left(x^2+(1-x)^2\right)
\ln\frac{W^2}{4m_l^2}+\left(x^2+(1-x)^2\right)\ln 4+8x(1-x)-1.
\label{C04}
\end{eqnarray}
The lepton mass regularizes mass singularities of (\ref{C04}), as
does the quark mass in (\ref{pureQED}). Since in the kinematically accessible
region $W^2\ge 4m_l^2$, corresponding to
$x\le 1/(1+4m_l^2/Q^2)$, both the function
$B(x)=(x^2+(1-x)^2)\ln 4+8x(1-x)-1$ and the first term in (\ref{C04}) are
positive, there is thus no problem with positivity of $F^{\gamma}_{lept}$.
The same holds for the QED
contribution to quark distribution function with $m_q$ as infrared regulator.

To identify the kinematical regions contributing to various parts
of (\ref{pureQED}) or (\ref{C04}) we return to the $x$-space, where the limits
on the virtuality $\tau$ (see Fig. \ref{figpl}) of the quark included in the
definition of the pointlike quark distribution function
of the photon are given as \cite{smarkem}
\be
\tau_{\mathrm{min}}=\frac{m_q^2}{1-x}\le \tau \le \frac{Q^2}{x}=
\tau_{\mathrm{max}}.
\label{bounds}
\ee
The potentially troubling term $\ln(1-x)$ in (\ref{C0}) comes from the lower
limit on $\tau$. So long as it appears in the combination $\ln(Q^2(1-x)/m_q^2x)$
as it does in (\ref{pureQED}) it is harmless as the whole term
$\ln(Q^2(1-x)/m_qx)$ stays, as argued above, positive. The problem, does,
however, arise when
we replace $m_q$ in this combination with an arbitrary initial $M_0$, which,
moreover, is typically around 1 GeV and consider (\ref{pureQED}) for
$x\ge 1/(1+4M_0^2/Q^2)$, where $\ln(Q^2(1-x)/M_0^2x)\le 0$. In this case we
should drop the term proportional to $\ln(1-x)$ which comes from the region
cut-off by the introduction of $M_0$. Moreover, we should also drop all other
terms which come predominantly from the
lower integration range in (\ref{bounds}). As argued in \cite{ja,smarkem} this
implies that $C_{\gamma}^{(0)}$ as given in (\ref{C0}) should be replaced with
\be
C_{\gamma}^{(0)}(x,Q/M)=\left[x^2+(1-x)^2\right]\ln\frac{Q^2}{M^2}+
\left[x^2+(1-x)^2\right]\ln \frac{1}{x}+6x(1-x)-1,
\label{new}
\ee
which implies decent behaviour of
\be
C_{\gamma}^{(0)}(x,1)=
\left[x^2+(1-x)^2\right]\ln \frac{1}{x}+6x(1-x)-1
\label{new2}
\ee
when $x\rightarrow 1$. Note that the expression (\ref{new2}) coincides
with that used in \cite{sas}.

\section{Summary and Conclusions}
The standard definition of LO and NLO approximations to photon structure function
are shown to be inconsistent with the requirement of factorization scale and scheme
invariance. The origin of this shortcoming is traced back to the usual but incorrect
claim that quark distribution functions of the photon behave as $\alpha/\alpha_s$.

Theoretically consistent definition of LO, NLO and NNLO approximations for the
photon structure function are constructed and the potentially troubling behaviour
of $C_{\gamma}^{(0)}(x,Q/M)$ at large $x$ is removed by discarding the terms coming
from the region outside the validity of perturbation theory.

The so called DIS$_{\gamma}$ factorization scheme invented in order to cure the
mentioned problem with large-$x$ behaviour of $C_{\gamma}^{(0)}(x,Q/M)$ is shown
to be ill-defined as it is based on the procedure which violates the factorization
scheme invariance of the photon structure function.

\vspace*{0.5cm}
\noindent
{\Large \bf Acknowledgment}

\vspace*{0.3cm}
\noindent
This work has been supported by the the project AV0-Z10100502 of the
Academy of Sciences of the Czech Republic and project LC527 of the Ministry
of Education of the Czech Republic. I am grateful to K. Kol\'{a}\v{r} for
careful reading of the manuscript.

\end{document}